\documentclass[preprint,floats,12pt]{article}
\usepackage{amssymb,amsmath}
\usepackage{eqsection,latexsym,epsf}
\usepackage{epsfig,amsfonts,cite}

\newenvironment{pseudocode}[1]{
\begin{enumerate}\tt\small\linewidth=11cm
\item[]\underline{{#1}}}
{\end{enumerate}}

\newcommand{\beq}{\begin{equation}}
\newcommand{\eeq}{\end{equation}}
\newcommand{\beqa}{\begin{eqnarray}}
\newcommand{\eeqa}{\end{eqnarray}}

\newcommand{\atanh}{{\rm atanh}}

\def\G{{\cal G}}
\def\E{{\cal E}}
\def\V{{\cal V}}
\def\A{{\cal A}}
\def\tC{\tilde{C}}
\def\d{\partial}
\def\sla{{\backslash}}
\def\<{\langle}
\def\>{\rangle}
\def\bordo{{\sf\bf B}}
\def\centro{{\sf\bf K}}

\def\T{T}
\def\de{{\rm d}}
\def\Q{Q}

\def\ed{\stackrel{{\rm d}}{=}}
\def\expect{{\mathbb E}}
\def\Z{{\mathbb Z}}
\def\bC{\overline{C}}
\def\p{\hat{p}}
\def\vs{\vec{\sigma}}

\title{How to Compute Loop Corrections to Bethe Approximation}

\author{
  { Andrea Montanari and Tommaso Rizzo}              \\
  {\small\it Laboratoire de Physique Th\'{e}orique de l'Ecole Normale
  Sup\'{e}rieure\footnote {UMR 8549, Unit{\'e}   Mixte de Recherche du
Centre National de la Recherche Scientifique et de
l' Ecole Normale Sup{\'e}rieure. } }
  \\[-0.2cm]
  {\small\it 24, rue Lhomond, 75231 Paris CEDEX 05, FRANCE}        \\[-0.2cm]
  {\small Internet: {\tt montanar@lpt.ens.fr, rizzo@lpt.ens.fr}}
          \\[-0.1cm]
  {\protect\makebox[5in]{\quad}}  % To force authors' names to be written
}

\date{}

\begin{document}

\maketitle

\thispagestyle{empty}

\abstract{We introduce a method for computing corrections to Bethe 
approximation for spin models on arbitrary lattices. Unlike
cluster variational methods, the new approach takes into account
fluctuations on all length scales. 

The derivation of the leading correction is explained and applied to two simple
examples: the ferromagnetic Ising model on $d$-dimensional lattices, and
the spin glass on random graphs (both in their high-temperature
phases). In the first case we rederive 
the well-known Ginzburg criterion and the upper critical dimension. In the 
second, we compute finite-size corrections to the free energy.}

\clearpage
%
%******************************************************
%
\section{Introduction}

Mean field approximations are the among the most frequently
used tools in Statistical Physics. Among them, Bethe approximation (BA)
\cite{Bethe} 
allows to treat with reasonable accuracy a large variety of lattice
models. Recently it has been successfully applied 
(in an algorithmic form) to problems of inference~\cite{Jordan,Pearl,Yedidia}, 
communications~\cite{RichardsonUrbanke}, and combinatorial 
optimization~\cite{MarcGiorgioRiccardo}. 
Often in these cases the underlying lattice has few or no short loops, and BA
(which is exact on trees) can become exact in the thermodynamic 
limit.

BA can be systematically improved using Kikuchi~\cite{Kikuchi}
or cluster variational methods (CVM). 
These approaches take into account `exactly'
of correlations up to some finite range $r$ and their complexity grows
exponentially with $r$. Because of this feature, they are 
unsuited for understanding the effect of long length scale fluctuations.
Furthermore, in lattices without short loops, no improvement is 
obtained unless $r$ is very large (which is of course unfeasible). 

For models on $d$-dimensional lattices, mean field can be also regarded 
as the zeroth order term in a $1/d$ expansion.\footnote{Generally, BA takes 
into account exactly also  the first $1/d$ correction.} Such an expansion is
however close in spirit to CVM, in that it keeps into account only the 
effect of short loops in the lattice~\cite{GeorgesYedidia}.
 
In the field-theoretical setting~\cite{ZinnJustin,Cardy}, mean field 
approximation is usually derived
by retaining only tree-level Feynman diagrams. The usual loop expansion
improves systematically over such an approximation by taking into account
of fluctuations on all length scales order-by-order in a properly defined 
coupling parameter. When resummed using renormalization-group ideas, it
gives an accurate description of many critical phenomena. 

Often a simple (and correct) field-theoretical formulation of the problem
is hard to derive. 
This is the case, for instance, of problems with quenched disorder,
where one usually invoke the replica trick for averaging over the 
disorder~\cite{Cardy}.
Also, field theoretical methods are usually unreliable for 
computing non-universal quantities. These can be on the other hand important 
for some of the applications (inference, communications, optimization) 
mentioned above. In this paper we present an approach 
for computing corrections to BA coming from fluctuations on all length scales.

To be concrete, we shall focus on  spin models with pairwise 
interactions on general graphs, with Hamiltonian
\begin{eqnarray}
E(\sigma) = -\sum_{(ij)\in\G}J_{ij}\sigma_i\sigma_j -
\sum_{i=1}^{N}H_i\sigma_i\, .\label{eq:Hamiltonian}
\end{eqnarray}
Here $\G = (\V,\E)$ is a graph with vertex set $\V= \{1,\dots,N\}$ and 
edges $\E \ni (i, j)$, $\E\subseteq\V\times\V$. 
The set of neighbors of the site $i$ is noted $\d i$.
We shall use the letters $a,b,\dots$ to
denote generic edges, and, whenever necessary write $(i_a,j_a) = a$. 
Finally, for any set of vertices $A\in\V$, 
$\sigma_A\equiv\{\sigma_i:i\in A\}$.  
Several families of graphs and choices of the couplings $J_{ij}, H_i$ 
will be considered in Section~\ref{se:Applications}, but for the time 
being we shall remain completely general. 

The Bethe approximation for such a model~\cite{MezardParisiBethe}
is better described by introducing a field $h_{i}^{(j)}$ 
for each directed link $i\to j$ of $\G$. Such fields are
required to satisfy the equations
\begin{eqnarray}
h_i^{(j)} = H_i + \sum_{l\in\d i\sla j} u_{J_{ij}}(h_{l}^{(i)})\, ,\;\;\;\;
\;\;\;
u_J(h) \equiv \frac{1}{\beta}\,\atanh[\tanh(\beta J)\tanh(\beta h)]\, .
\label{eq:BetheEquations}
\end{eqnarray}
Once a solution of these equations is found,  one can use the fields 
$h_{i}^{(j)}$ to estimate the thermal average of local operators. For instance
\begin{eqnarray}
\<\sigma_i\> & \stackrel{\rm Bethe}{=} &\frac{1}{w_i} 
\sum_{\sigma=\pm 1}\sigma\, \exp(\beta H_i\sigma)
\prod_{j\in\d i}\frac{e^{\beta u(J_{ij},h_{j}^{(i)})\sigma}}
{2\cosh(\beta u(J_{ij},h_{j}^{(i)}))}
\, ,\label{eq:BetheExpectationValue}\\
w_i & = & \sum_{\sigma=\pm 1}\exp(\beta H_i\sigma)
\prod_{j\in\d i}\frac{e^{\beta u(J_{ij},h_{j}^{(i)})\sigma}}
{2\cosh(\beta u(J_{ij},h_{j}^{(i)}))}
\, .
\end{eqnarray}
The basic approximation involved in derived these equations is the following.
Consider a spin $\sigma_{i}$ and set its interaction with the
neighbors to $0$: $J_{ij} = 0$ for all $j\in\partial i$ (in other words
$\sigma_i$ is `removed' from the system). Now look at the joint probability 
distribution of the neighboring spins $\sigma_{\partial i}$ in the system without $\sigma_i$.
Bethe approximation amount to saying that 
\begin{eqnarray}
P_i(\sigma_{\partial i})\stackrel{\rm Bethe}{=} \prod_{j\in\partial i}
\frac{e^{\beta h_{j}^{(i)}\sigma_j}}{2\cosh\beta h_{j}^{(i)}}, \, .
\label{eq:BetheCavity}
\end{eqnarray}
Our approach consists in deriving a set of exact equations 
for the `cavity' distributions $P_i(\,\cdot\,)$'s. When the form 
(\ref{eq:BetheCavity}) is plugged in these equations, the Bethe
equations (\ref{eq:BetheEquations}) are derived. Corrections are computed 
by introducing correlations in  $P_{i}(\sigma_{\partial i})$.

In Section \ref{se:GeneralSection} we shall explain the general method 
for computing corrections to BA. We then use it in 
Section \ref{se:Applications} for computing the leading  corrections to 
BA for two particular examples: the ferromagnet on cubic $d$-dimensional 
lattices and the spin glass on random graphs.
%
%************************************************************************
%
\section{The general approach}
\label{se:GeneralSection}

\begin{figure}[t]
\begin{center}
\epsfig{file=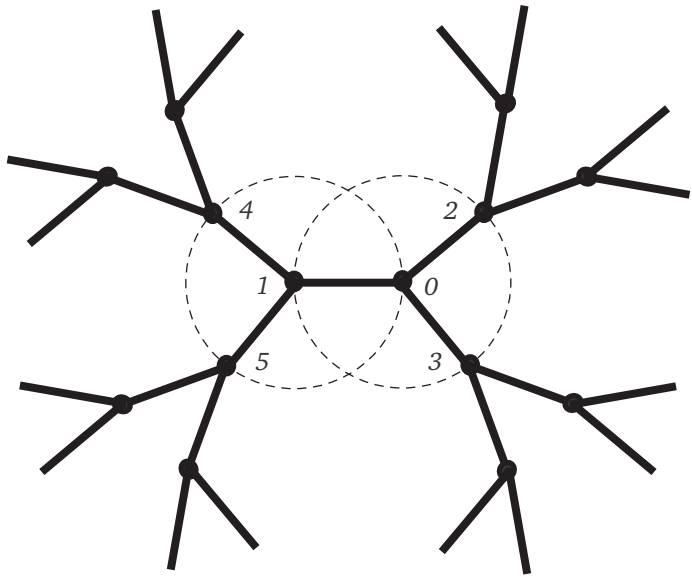} 
\caption{The magnetizations at site $0$ and $1$ in the absence of link $J_{10}$ can be expressed in term of the correlations of the spins $0,4,5$ in the absence of spin $1$ or of the correlations between spins $1,2,3$ in the absence of spin $0$, the equality of the results yields the cavity equations.}\vspace{-1.cm}
\label{fig:Conn3}
\end{center}
\end{figure}

In order to explain the general computation scheme, 
it is convenient to introduce some notation.
We denote by $E^{(i)}(\sigma)$ a modified energy function in which the
interactions between the spin $i$ and its neighbors have been canceled.
Analogously, $E^{(a)}(\sigma)$, with $a\in\E$, is the energy function 
modified by eliminating the interaction along the edge $a$.
In formulae:
\begin{eqnarray}
E^{(i)}(\sigma) = E(\sigma) +\sum_{j\in\d i}  J_{ij}\sigma_i\sigma_j\, ,
\;\;\;\;\;\;\; E^{(a)}(\sigma) = E(\sigma) + J_{i_aj_a}\sigma_{i_a}\sigma_{j_a}
\, .
\end{eqnarray}
We denote by $\<\,\cdot\,\>^{(i)}$ and $\<\,\cdot\,\>^{(a)}$ the Boltzmann averages
with respect to these modified energy functions. As in the 
introduction, $P_i(\sigma_{\d i})$ the marginal distribution  
of the neighbors of $i$ with respect to the system with
energy $E^{(i)}(\sigma)$. Analogously, we define the distribution
$P_a(\sigma_{i_a},\sigma_{j_a})$.

In order to have a concrete representation for
the distributions $P_{i}(\sigma_{\partial i})$, we shall use the correlation functions
\begin{eqnarray}
\tC^{(i)}_{\A} \equiv \<\prod_{j\in\A}\sigma_j\>^{(i)} = \sum_{\sigma_{\d i}}
P_i(\sigma_{\d i})\; \prod_{j\in\A}\sigma_j\, .
\end{eqnarray}
for any non-empty subset $\A\in\d i$. In the special case $\A = \{j\}$
we shall also use the more conventional notation $M^{(i)}_j = \tC^{(i)}_{j}$.
In  Bethe approximation, the distribution $P_i(\sigma_{\d i})$ is
assumed to be factorized, cf. Eq.~(\ref{eq:BetheCavity}). 
In order to compute corrections, it is convenient to introduce the connected 
correlation functions $C^{(i)}_{\A}$. We have the usual relation
\begin{eqnarray}
\tC^{(i)}_{\A} = \sum_{[\A_1\dots \A_n]} 
C^{(i)}_{\A_1}\cdots C^{(i)}_{\A_n}\, ,\label{eq:Connected}
\end{eqnarray}
with $[\A_1,\dots,\A_n]$ running over the partitions of $\A$. 
Finally $C^{(i)}\equiv \{C^{(i)}_\A\, :\, \A\subseteq \d i\}$.

Let us now derive the basic relation between the $P_i(\,\cdot\,)$'s
to be exploited in the following.
Consider two sites $i$ and $j$ which are joined by an edge in
$\G$. We can construct the distribution $P_{(ij)}(\sigma_i,\sigma_j)$
in two ways:
\begin{eqnarray}
P_{(ij)}(\sigma_i,\sigma_j) = \frac{1}{Z_{j}}\sum_{\sigma_{\d j\sla i}}
P_j(\sigma_{\d j}) \, \exp\Big\{\beta H_j\sigma_j+\beta\sum_{l\in\d j\sla i}
J_{jl}\sigma_j\sigma_l\Big\}\, ,\label{eq:LinkDistr}
\end{eqnarray}
and the equivalent one (let us call $P'_{ij}(\sigma_i,\sigma_j)$ the
corresponding expression) which is obtained by interchanging $i$ and $j$.
Here $Z_j$ is a constant which ensures the correct normalization of
$P_{(ij)}(\sigma_i,\sigma_j)$. We can now marginalize the
right hand side of Eq.~(\ref{eq:LinkDistr}) with respect to 
$\sigma_j$ (to $\sigma_i$) in order to compute the magnetizations
on site $i$ (site $j$) with respect to the system with 
energy function $E^{(ij)}(\sigma)$. The same calculation can be performed
using the expression $P'_{(ij)}(\sigma_i,\sigma_j)$. Since the result 
of these two calculations must be the same, we obtain two equations
of the form:
\begin{eqnarray}
\bordo^{(i)}_j ( C^{(i)} ) = \centro^{(j)}_i( C^{(j)})\, ,\;\;\;\;\;\;\;
\centro^{(i)}_j (C^{(i)}) = \bordo^{(j)}_i(C^{(j)})\, .
\label{eq:BordoCentro}
\end{eqnarray}
The function $\bordo^{(i)}_j(\, \cdot\, )$ yields the magnetization at site
$j$ when the distribution $P_i(\sigma_{\d i})$ is modified through 
the addition of the interactions $J_{il}$, $l\neq j$.
Analogously, $\centro^{(i)}_j(\, \cdot\, )$ yields the magnetization at site 
$i$ for the same system. Elementary algebraic manipulations 
yields the explicit expressions:
\begin{eqnarray}
\bordo^{(i)}_j ( C ) & = & \frac{\underset{\A \, \mbox{\tiny even}}{\sum} t_\A\tC_{\A\cup j}+t(H_i)
\underset{\A \, \mbox{\tiny odd}}{\sum}   t_\A\tC_{\A\cup j}}
{\underset{\A \, \mbox{\tiny even}}{\sum} t_\A\tC_{\A}+t(H_i)
\underset{\A \, \mbox{\tiny odd}}{\sum} t_\A\tC_{\A}}\, ,
\label{eq:BordoDef}\\
\centro^{(i)}_j (C)  & = & 
\frac{t(H_i)\underset{\A \, \mbox{\tiny even}}{\sum} t_\A\tC_{\A}+
\underset{\A \, \mbox{\tiny odd}}{\sum}   t_\A\tC_{\A}}
{\underset{\A \, \mbox{\tiny even}}{\sum} t_\A\tC_{\A}+t(H_i)
\underset{\A \, \mbox{\tiny odd}}{\sum} t_\A\tC_{\A}}\, ,
\label{eq:CentroDef}
\end{eqnarray}
where the sums over $\A$ run over all the subsets of neighbors of
$i$, $\A\in\d i$, which do not include $j$. Furthermore, we used the
shorthands $t_{\A} \equiv \prod_{l\in\A}t_{il}$, 
$t_{il}\equiv \tanh(\beta J_{il})$ and $t(H_i)\equiv
\tanh(\beta H_i)$. The above functions can be 
written in terms of the connected correlations $C^{(i)}_{\A}$ by using the 
relation (\ref{eq:Connected}).

Consider for instance the case depicted in Fig.~\ref{fig:Conn3} 
where $i=0$ and 
$j=1$ have both degree 3. We furthermore assume, for the sake of 
simplicity, $H_0 = H_1 = 0$. Then
\begin{eqnarray}
\bordo_1^{(0)}(C^{(0)}) &=&  M_1^{(0)}+t_{02}t_{03}\,
\frac{C_{21}^{(0)}M_{3}^{(0)}+C_{31}^{(0)}
M_{2}^{(0)}+C_{123}^{(0)}}{1+t_{02}t_{03}M_{2}^{(0)}M_{3}^{(0)}+
t_{02}t_{03}C_{23}^{(0)}}\, ,\\
\centro_1^{(0)}(C^{(0)})  &=& \frac{t_{02}M_{2}^{(0)}+t_{03}M_{3}^{(0)}}{
    1+t_{02}t_{03}M_{2}^{(0)}M_{3}^{(0)}+t_{02}t_{03}C_{23}^{(0)}}\, .
\end{eqnarray}
The analogous expressions for $\bordo_0^{(1)}(C^{(1)})$ and 
$\centro_0^{(1)}(C^{(1)})$  
are obtained by interchanging $0\leftrightarrow 1$ and $\{2,3\}\leftrightarrow\{4,5\}$.
It is therefore easy to write explicitly the equation 
$\bordo^{(1)}_0(C^{(1)}) = \centro^{(0)}_1(C^{(0)})$:
\begin{eqnarray}
M_{0}^{(1)}+t_{14}t_{15}\,\frac{C_{40}^{(1)}M_{5}^{(1)}+C_{50}^{(1)}M_{4}^{(1)}+C_{045}^{(1)}}{1+t_{14}t_{15}M_{4}^{(1)}M_{5}^{(1)}+t_{14}t_{15}C_{45}^{(1)}}
=  \frac{t_{02}M_{2}^{(0)}+t_{03}M_{3}^{(0)}}
{1+t_{02}t_{03}M_{2}^{(0)}M_{3}^{(0)}+t_{02}t_{03}C_{23}^{(0)}}
\nonumber\, .
\end{eqnarray}

There are $2|\E|$ equations of the form (\ref{eq:BordoCentro}):
one for each directed link in the graph. The number of unknowns is,
on the other hand $\sum_{i} (2^{|\d i|}-1)$, with the sum running
over the sites of the graph, and $|\d i|$ being their connectivity.
Therefore, these equations are not sufficient to determine the
correlation functions $C^{(i)}$. If, on the other hand, we neglect multi-spin
connected correlation functions and only retain the cavity magnetizations 
$M^{(i)}_j$ (as in Bethe approximation), we are left with 
$2|\E|$ variables to determine. In this case Eqs.~(\ref{eq:BordoDef}) 
and (\ref{eq:CentroDef}) are considerably simplified:
\begin{eqnarray}
\bordo^{(i)}_j ( C^{(i)} ) & \stackrel{\rm Bethe}{=} & M_{j}^{(i)}\, , \\
\centro^{(i)}_j( C^{(i)} ) & \stackrel{\rm Bethe}{=} &
\tanh\left\{\beta H_i+
\sum_{k\in\partial i\backslash j}\atanh\left[ t_{ik}M^{(i)}_{k}\right]\right\}\, .
\end{eqnarray}
By setting $M^{(i)}_j = \tanh\beta h^{(i)}_j$, it is easy to see that the equations
(\ref{eq:BordoCentro}) are in this case the Bethe 
equations\footnote{This derivation of BA is in fact mentioned as a side remark 
in Ref.~\cite{Yedidia2}.} (\ref{eq:BetheEquations}).
If, for instance, $\G$ is a tree, the connected cavity 
correlations $C^{(i)}_{\A}$ vanish if $|\A|\ge 2$.  We thus proved 
recovered the well-known result that Bethe approximation is exact
on tree graphs.

We want now to estimate the connected cavity correlations $C^{(i)}_{\A}$, 
for $|\A|\ge 2$, and then use Eq.~(\ref{eq:BordoCentro}) to improve 
the calculation of $M^{(i)}_j$. In synthesis, the correlations are estimated
through the fluctuation-dissipation theorem:
\begin{eqnarray}
\frac{1}{\beta^{n}}\frac{\partial^{n} M_{i_1}}{\partial H_{i_2}\cdots
\partial H_{i_{n+1}}} = C_{i_1\cdots i_{n+1}}\, ,
\label{eq:FluctuationDissipation}
\end{eqnarray}
where $i_1\dots i_{n+1}$ are $n+1$ distinct index sites. Here 
$M_i$ and $C_{i_1\cdots i_{n+1}}$ are magnetizations and correlations 
with respect to an arbitrary Hamiltonian of the form
(\ref{eq:Hamiltonian}). In other terms, one can obtain 
equations for the correlations by taking appropriate derivatives 
of the exact equations (\ref{eq:BordoCentro}) with respect to an external 
field.\footnote{One can see that the differentiation procedure is 
well-defined through the following (numerically imprecise but 
conceptually simple) implementation. Compute the magnetization 
$M_i$, then change slightly the external field on site $j$, 
$H_j\to H_j+\delta H_j$, and evaluate the correlation function
between sites $i$ and $j$ as $C_{ij} = \lim_{\delta H_j\to 0}
(\delta M_i/\beta \delta H_j)$. Higher-order correlations are computed 
analogously by considering variations of the external fields at 
several points.}

For, the sake of clarity, let us compute the leading-order correction to
BA. We neglect all connected cavity correlation
functions $C^{(i)}_{\A}$ with $|\A|\ge 3$. Moreover, we treat two point
correlation functions to the linear order. To this order, the expressions 
(\ref{eq:BordoDef}) and (\ref{eq:CentroDef}) become 
\begin{eqnarray}
\bordo^{(i)}_j ( C ) & = & M^{(i)}_j + \sum_{l\in\partial i\backslash j}
\Omega^{(i)}_{j,l}\,t_{il}
C^{(i)}_{jl}+O(C^2)\, ,\label{eq:ExpBordo}\\
\centro^{(i)}_j( C ) & = & \T^{(i)}_j + \sum_{(l_1,l_2)\in\partial i\backslash
j}\Gamma^{(i)}_{j,l_1l_2} t_{il_1}t_{il_2}C^{(i)}_{l_1l_2}+O(C^2)\, ,
\label{eq:ExpCentro}
\end{eqnarray}
where
\begin{eqnarray}
\Omega^{(i)}_{j,l} &= &\frac{\T^{(i)}_{jl}}{1+t_{il}M^{(i)}_l\T^{(i)}_{jl}}\, ,
\label{eq:OmegaDef}\\
\Gamma^{(i)}_{j,l_1l_2} &=&\frac{\T^{(i)}_{jl_1l_2}-\T^{(i)}_j }
{1+t_{il_1}t_{il_2}M^{(i)}_{l_1}M^{(i)}_{l_2}+
t_{il_1}M^{(i)}_{l_1}\T^{(i)}_{jl_1l_2}+t_{il_2}M^{(i)}_{l_2}
\T^{(i)}_{jl_1l_2}} \, ,\label{eq:GammaDef}
\end{eqnarray}
and
\begin{eqnarray}
\T^{(i)}_{l_1l_2\dots} & = &\tanh\left\{\beta H_i+
\sum_{k\in\partial i\backslash l_1,l_2\dots}
\atanh\left[ t_{ik}M^{(i)}_{k}\right]\right\}\, .
\end{eqnarray}
We can therefore proceed as follows. 
First solve the Bethe equations (\ref{eq:BetheEquations})
for the original energy function (\ref{eq:Hamiltonian}). Then,
for each site $i\in\V$, consider the energy function $E^{(i)}(\sigma)$ 
corresponding to the spin $\sigma_{i}$ being removed. Compute 
the two point connected correlation functions in the reduced system
$C^{(i)}_{j_1j_2}$ using  BA together with the fluctuation-dissipation
relations (\ref{eq:FluctuationDissipation}). Finally write
\begin{eqnarray}
M^{(i)}_j = \tanh(\beta h^{(i)}_j) + \Delta M^{(i)}_j + O(C^2)\, .
\end{eqnarray}
The first order corrections $\Delta M^{(i)}_j$ are computed by expanding 
the equations (\ref{eq:BordoCentro}) up to first order in $C$.
Using the expansion (\ref{eq:ExpBordo}), (\ref{eq:ExpBordo}), we get 
\begin{eqnarray}
\Delta M^{(i)}_j+\sum_{l\in\partial i\backslash j}t_{il}
\Omega^{(i)}_{j,l}C^{(i)}_{jl} =\! \sum_{k\in\partial j\backslash i}
\Q^{(j)}_{i,k}\, \Delta M^{(j)}_k
+\sum_{(l_1,l_2)\in\partial j\backslash
i}\!\!\Gamma^{(j)}_{i,l_1l_2} t_{jl_1}t_{jl_2}C^{(j)}_{l_1l_2}\, .
\label{eq:CorrectionEq}
\end{eqnarray}
where
\begin{eqnarray}
\Q^{(j)}_{i,k} =
\frac{t_{jk} [1-(T^{(j)}_i)^2]}{1-(t_{jk}\tanh(\beta h^{(j)}_k))^2}\, \Delta M^{(j)}_k
\end{eqnarray}
Here the coefficients $\Omega^{(i)}_{j,l}$ and $\Gamma^{(j)}_{i,l_1l_2}$
are computed through Eqs~(\ref{eq:OmegaDef}) and (\ref{eq:GammaDef})
by setting $M^{(l)}_k\to\tanh(\beta h^{(l)}_k)$.
We thus obtained one equation of the form (\ref{eq:CorrectionEq}) for each 
directed edge in the lattice. These completely determine the $\Delta 
M^{(i)}_j$.

\begin{figure}
\begin{center}
\epsfig{file=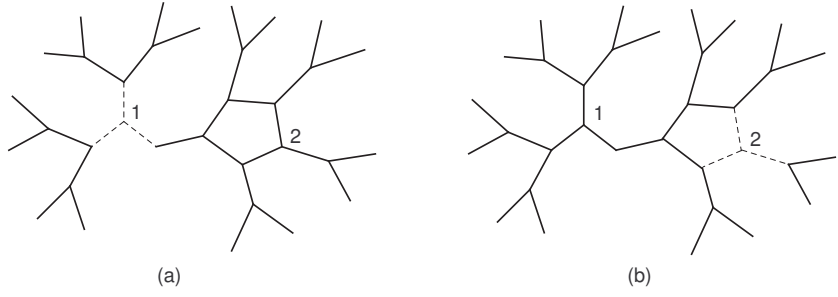,angle=0,width=0.7\linewidth} 
\caption{Graph $\G$ with a single loop. The first-order procedure
described in the text is exact on such a model. This claim can be proved
by considering two type cavities (a) and (b).}\vspace{-1.cm}
\label{fig:OneLoop}
\end{center}
\end{figure}
It is important to stress that the above procedure is not uniquely
defined. One could consider equivalent calculations which differ 
from the above by higher order corrections. Here are two examples:
\begin{enumerate}
\item One could compute the connected cavity correlations without 
removing the spin $i$ from the system but rather using
the relation 
\begin{eqnarray}
C^{(i)}_{jl} = (1-\tanh^2\beta h^{(i)}_j)\, \beta\frac{\partial h^{(i)}_j}
{\partial H_l}\, ,\label{eq:FluctuationDissipationSpecial}
\end{eqnarray}
which in turns follows from $M^{(i)}_j = \tanh \beta h^{(i)}_j$. 
Equations for $\frac{\partial h^{(i)}_j}{\partial H_l}$ are easily obtained by 
differentiating Eq.~(\ref{eq:BetheEquations}).
\item Instead of expanding the equations (\ref{eq:BordoCentro})
to the first order in $C^{(i)}_{jl}$ and $\Delta M^{(i)}_j$,
one could proceed as follows. Set to zero all the correlation functions
$C^{(i)}_{\A}$, $|\A|\ge 3$, replace $C^{(i)}_{jl}$ by their 
Bethe approximation, and solve for $M^{(i)}_j$.
\end{enumerate}
In particular, the last implementation is exact
whenever the graph $\G$ contains (at most)  a unique loop
(thus improving over BA). Since the equations (\ref{eq:BordoCentro}) 
are exact, in order to prove this claim 
it is enough to show that the procedure defined above 
computes correctly the correlation functions $C^{(i)}_{\A}$, $|\A|\ge 2$.
If $|\A|\ge 3$ and $\G$ has a single loop, then $C^{(i)}_{\A}=0$,
and the algorithm does not make any error in neglecting these correlations.
Consider now the computation of $C^{(i)}_{l_1 l_2}$ and distinguish two cases.
In the first case, the graph obtained by removing the vertex $i$
is a tree (as e.g. for site $1$ in Fig.~\ref{fig:OneLoop}). 
Therefore BA is exact for the energy function 
$E^{(i)}(\sigma)$ and correctly computes the correlations $C^{(i)}_{l_1 l_2}$.
In the second case upon removing site $i$, the resulting graph,
let us call it $\G\backslash i$, still 
contains a loop (as for site $2$ in Fig.~\ref{fig:OneLoop}). 
Notice that $\G\backslash i$ is disconnected, and each of the neighbors of $i$
belongs to a distinct connected component. Therefore $C^{(i)}_{j_1j_2}=0$
for any two neighbors $j_1,j_2$ of $i$. While BA is not exact for 
$\G\backslash i$,
it is easy to see that it correctly gives yields vanishing correlations among 
sites belonging to different connected components.

At this point it should be clear how to improve the above 
first-order scheme. After removing the spin $i$, one can compute the 
correlations $C^{(i)}_{\A}$ within the first order scheme rather than Bethe
approximation (and retain three points correlation as well) and then 
recompute the magnetizations $M^{(i)}_j$ using Eq.~(\ref{eq:BordoCentro}).

The general procedure can be explained as a recursive 
pseudocode. The code makes use of a routine 
{\tt Correlation(~$\G$,~$E(\cdot)$,~$\{ C^{(i)}\}$~)} 
which takes as  input a graph $\G$, an Hamiltonian $E(\cdot)$ 
of the form (\ref{eq:Hamiltonian}), an estimation of the $n$-points
cavity correlations $\{ C^{(i)}\}$ for any $i\in\G$.
The output  consists of a new estimation of all $n$-points correlation 
functions. This is obtained
(let's repeat ourselves) by a joint solution of Eqs.~(\ref{eq:BordoCentro})
and (\ref{eq:FluctuationDissipation}). 
A particular case of the routine  {\tt Correlation(~$\cdot$~)}
is obtained when all the multi-point cavity correlations 
$\{ C^{(i)}\}$ are set to $0$.
This corresponds of course to  BA. 
For the sake of clarity, we shall denote
the corresponding routine {\tt Bethe(~$\G$,~$E(\cdot)$~)}
instead of {\tt Correlation(~$\G$,~$E(\cdot)$,~$0$~)}.

The recursive routine, ${\tt Loop()}$, 
takes as input a graph $\G$, an Hamiltonian of the form 
(\ref{eq:Hamiltonian}), and the order of approximation $\ell$ to be 
achieved. The output consists of an estimation of all 
$n$-points connected correlation functions in the system.
\begin{pseudocode}{Loop( $\G$, $E(\cdot)$, $\ell$ )}
\item[] If ($\ell == 1$) Output Bethe( $\G$, $E(\cdot)$ )
\item[] Else
\begin{itemize}
\item[] For ($i\in \V$) $C^{(i)} : =$ 
Loop( $\G\sla i$, $E^{(i)}(\cdot)$, $\ell-1$ )
\item[] Output Correlation( ~$\G$,~$E(\cdot)$,~$\{ C^{(i)}\}$~)
\end{itemize}
\item[] End
\end{pseudocode}
Let us stress that this algorithm  deals with particular samples of 
the model, without need for an average over disorder realizations.
Its complexity is (for graphs with bounded connectivity) $O(N^{\ell})$,
i.e. polynomial for any fixed $\ell$.
This makes its application to inference/optimization problems a viable
research direction. The algorithm implements the strategy 2 above
(actual elimination of a site): arguing as above, it can be proved that 
by induction that {\tt Loop( $\cdot$, $\cdot$, $\ell$ )} is exact on 
graphs with cyclic number not larger than $\ell$.
%
%************************************************************************
%
\section{Applications}
\label{se:Applications}

In this Section we apply the method developed so far to two simple problems:
the spin glass on random graphs with general connectivity distributions, 
and the ferromagnetic Ising model on the cubic $d$-dimensional lattice.
In both examples we will keep ourselves to the high temperature,
 no external field phase.  The precise nature of the corrections
computed within our approach is different for these two applications. In the
first case, they correspond to higher orders in the loop expansion.
While there is no formal parameter to order the loop expansion for
the Ising model on cubic lattices, we are able to recover the well known
Ginzburg criterion and the associated one-loop integral.
In the second one, successive terms in our expansion correspond
to higher powers in $1/N$  ($N$ being the number of spins).
%
%***********************************************
%
\subsection{Ising model on the cubic lattice}

We take $\G$ to be the $d$--dimensional cubic lattice, i.e. $\Z^d$ with 
edges joining neighboring vertices. 
Vertices of the lattices will be denoted in this Section
by $x,y,z,\dots \in \Z^d$, while unit vectors by 
$\mu,\nu,\dots\in\{(1,0,\dots), (0,1,\dots),\dots\}$.
We consider the usual
Ising ferromagnet, i.e. $J_{xy} = 1$ and $H_{x} = H$.

Because of translational invariance, the Bethe equations admit an 
uniform solution $h^{(x)}_y = h$, with $h$ solving 
\begin{eqnarray}
h = H + \frac{(2d-1)}{\beta}\, \atanh[\tanh\beta\tanh\beta h]\, .
\end{eqnarray}
Analogously, the cavity magnetizations do not depend on the site:
$M^{(i)}_j = M^{\rm cav}$, and we have $M^{\rm cav} = \tanh \beta h
+\Delta M^{\rm cav} + O(C^2)$. 
In order to write the results in a compact form, it is convenient to 
introduce the field
\begin{eqnarray}
h_{p} \equiv H+\frac{p}{\beta}\,\atanh[\tanh\beta\tanh\beta h]\, .
\end{eqnarray}
We shall furthermore use the shorthand $t \equiv \tanh\beta$.
After some tedious but straightforward 
calculations, Eq.~(\ref{eq:CorrectionEq}) yields
\begin{eqnarray}
\Delta M^{\rm cav} = -\frac{A}{B}\sum_{(\mu,\nu)}C^{(0)}_{\mu,\nu}\, ,
\end{eqnarray}
where 
\begin{eqnarray}
A & = &\frac{1}{d}\, \frac{t\,\tanh\beta h_{2d-2}}
{(1+t\tanh\beta h\tanh\beta h_{2d-2})(1-\tanh^2\beta h)}
+\frac{d-1}{d}\,\frac{2t^3\tanh\beta h}{1-t^2\tanh^2\beta h}
\, ,\phantom{AA}\\
B & = & \frac{1}{1-\tanh^2\beta h}-(2d-1)\, \frac{t}{1-t^2\tanh^2\beta h}\, .
\end{eqnarray}

Using these expressions, we can compute the (non-cavity) magnetization
$M =\<\sigma_x\>$
\begin{eqnarray}
M & =&M_0 + M_1\sum_{(\mu,\nu)}C^{(0)}_{\mu,\nu}+O(C^2)\, ,\\ 
M_0 & = & \tanh\beta h_{2d}\, ,\\ 
M_1 & = & t^2\, \frac{\tanh\beta h_{2d-2}-
\tanh\beta h_{2d}}{1+t^2\tanh^2\beta h+2t\tanh\beta h\tanh\beta h_{2d-2}}
-2dt \, \frac{1-\tanh^2\beta h_{2d}}{1-t^2\tanh^2\beta h}\, \frac{A}{B}\, .
\phantom{AA}
\end{eqnarray}
By taking the $H\to 0$ limit, we can compute the zero field susceptibility
defined by $M = \chi H+O(H^2)$. This admits the expansion
$\chi = \chi_0 + \chi_1\sum_{(\mu,\nu)}C^{(0)}_{\mu,\nu}+O(C^2)$,
where
\begin{eqnarray}
\chi_0  = \beta\, \frac{1+t}{1-(2d-1)t}\, ,\;\;\;\;\;\;\;\;
\chi_1  = -2\, \beta\,\frac{t^2(1-t^2)}{[1-(2d-1)t]^2}\, .\label{eq:ChiExp}
\end{eqnarray}

We are left with the task of computing the parameter 
$\bC \equiv \sum_{(\mu,\nu)}C^{(0)}_{\mu,\nu}$. We will follow the strategy
1 described in the previous Section. By differentiating the
Bethe equations, we get
\begin{eqnarray}
\frac{\partial h^{(x+\mu)}_x}{\partial H_z} = \delta_{x,z}+
t\, \frac{1-\tanh^2\beta h}{1-t^2\tanh^2\beta h}\, \sum_{\nu(\neq \mu)}
\frac{\partial h^{(x)}_{x+\nu}}{\partial H_z}\, .
\end{eqnarray}
Using the fluctuation-dissipation relation 
(\ref{eq:FluctuationDissipationSpecial}) and translation invariance, we get
the following equation for the cavity correlations
\begin{eqnarray}
&&C^{(0)}_{z,-\mu} = q\delta_{z,-\mu}+r \sum_{\nu(\neq \mu)}
C^{(0)}_{z+\mu,\nu}
\, ,\\
&&q\equiv 1-\tanh^2\beta h\, ,\;\;\;\;\;\;
r\equiv\frac{t}{1-t^2\tanh^2\beta h}\, .
\end{eqnarray}
These equations are easily solved by introducing the Fourier transform
\begin{eqnarray}
C^{(0)}_{\mu}(p) = \sum_{x}e^{ipx}C^{(0)}_{x,\mu}\, ,\;\;\;\;\;\;\;
C^{(0)}_{x,\mu} =\int\!\frac{\de^d p}{(2\pi)^d}\, C_{\mu}(p)\, e^{-ipx}\, ,
\end{eqnarray}
where the integral over $p$ runs over the Brillouin zone $[-\pi,\pi]^d$.
We obtain 
\begin{eqnarray}
C_{\mu}(p) = \frac{q(e^{ip\mu}-r)}{1-2dr+(2d-1)r^2+r\p^2}\, ,
\end{eqnarray}
where $\p^2\equiv 2d-\sum_{\mu} e^{ip\mu}$. Therefore the correlation parameter
entering in the corrections to BA is 
\begin{eqnarray}
\bC = -\frac{q}{2r^2} +\frac{(1-r^2)q}{2r^2}\int\!\frac{\de^d p}{(2\pi)^d}\,
\frac{1}{1-2dr+(2d-1)r^2+r\p^2}\, .
\end{eqnarray}
\begin{table}
\center{
\begin{tabular}{cccc}
\hline
$d$ & $\beta_{\rm c}^{\rm num}$ & $\beta_{\rm c}^{\rm new}$ &
$\beta_{\rm c}^{1/d}$\\
\hline
\hline
$3$ & $0.221654(6)$ \cite{Barber} & $0.237708$ & $0.207407$ \\
$4$ & $0.1497^\dagger$ \cite{KimPatrascioiu}& $0.151650$ & $0.145833$\\
$5$ & $0.11388(3)$ \cite{ParisiRuiz} & $0.114356$ & $0.112592$\\
$6$ &  $\ast$ & $0.092446$ & $0.091750$\\
\hline
\end{tabular}
}
\caption{Critical temperature for the Ising model on $d$--dimensional 
cubic lattices as determined by numerical simulations, the new expansion 
presented in this paper, and second order $1/d$ expansion.
The numerical result for $d=4$ is quoted in Ref.~\cite{KimPatrascioiu}
without statistical errors. A more accurate estimate 
$\beta_{\rm c} = 0.14966(3)$ was obtained from high temperature
expansion in Ref.~\cite{Gaunt}.}
\label{tab:CriticalT}
\end{table}
A simple application of the above calculation consists in computing 
the critical temperature. We can do this by solving the equation 
$\chi^{-1} = 0$, which to this order implies 
\begin{eqnarray}
1-(2d-1)\tanh^2\beta_{\rm c}+2\tanh^2\beta_{\rm c}(1-\tanh\beta_{\rm c})\bC =0\, ,
\end{eqnarray}
By solving this equation to the first order in $\bC$, we get
\begin{eqnarray}
\tanh \beta_{\rm c} =  \frac{1}{2d-1}-\frac{2d-2}{(2d-1)^3}[(2d-1)-2d(2d-2)I_d]\,,\\
I_d\equiv \int\!\frac{\de^d p}{(2\pi)^d}\, \frac{1}{\p^2}=
\int_{0}^{\infty} [e^{-2t}I_0(2t)]^d\, ,
\end{eqnarray}
where $I_0(\, \cdot\, )$ is the Bessel function.
The integral $I_d$ is convergent for $d> 2$.
A numerical calculation of $I_d$ yields the values
of $\beta_{\rm c}$ reported in Table \ref{tab:CriticalT}. This is compared 
with numerical simulations and with the second order $1/d$ 
expansion
\begin{eqnarray}
\beta^{1/d}_{\rm c} = \frac{1}{2d-1}\,\left[1+\frac{1}{3d^2}+O(d^{-3})
\right]\, .
\end{eqnarray}
We can also derive the critical behavior of the susceptibility. From
Eq.~(\ref{eq:ChiExp}) we get 
$\beta\chi^{-1} = K\, (\beta_{\rm c}-\beta) + O((\beta_{\rm c}-\beta)^2)$
where
\begin{eqnarray}
K = 2d-\left\{\frac{2d}{2d-1}-\frac{4d(2d-2)(2d^2-d+1)}{(2d-1)^2}\, I_d
+\frac{4d^2(2d-2)^2}{(2d-1)^2}\, J_d\right\}\, ,\\
J_d\equiv \int\!\frac{\de^d p}{(2\pi)^d}\, \frac{1}{(\p^2)^2}=
\int_{0}^{\infty}\! t\,[e^{-2t}I_0(2t)]^d\, .
\end{eqnarray}
The integral $J_d$ is infrared divergent for $d\le 4$. We have therefore
rederived the well known upper-critical dimension $d_{\rm up}=4$.
% 
%***********************************************
%
\subsection{Spin glass on random graphs}

We consider here the case in which $\G$ is an
Erd\"os-Renyi random graph with $N$ vertices and average connectivity
$\gamma$. Such a graph is constructed by drawing an edge between any couple 
$(i,j)$ of distinct vertices independently with probability $\gamma/N$. 
Spins joined by an edge interact via a coupling $J_{ij}$ which are
i.i.d. symmetric random variables with probability density function
$p(J)=p(-J)$. This is also known as the Viana-Bray model and was first 
studied in Ref.~\cite{VianaBray}. 
In the following $\expect$ will denote expectation with respect
to the couplings and/or the graph realization.
Finally we shall focus on the case of vanishing external field $H_i=0$.

The interest of such a simple model is that it can be easily
treated by the replica method, thus providing an useful check
of our approach.\footnote{Some of the properties of the high temperature
phase can be furthermore derived rigorously, see \cite{GuerraToninelli}.
For ideas on $1/N$ corrections in the low temperature phase see
\cite{Campellone}.} 
Before applying the strategy outlined in the 
previous Section, let us briefly recall the replica results.
Averaging over disorder, one gets the following representation for the 
moments of the partition function~\cite{MonassonDiluted}
\begin{eqnarray}
\expect\, Z^{n} = \int\exp\{-NS[c]\}\; {\cal D}c\, ,
\end{eqnarray}
where $c(\vs) = c(\sigma^1,\dots,\sigma^n)$, $\sigma^a\in\{\pm 1\}$
is the replica  order parameter.
The integral is then evaluated using the saddle point method,
the paramagnetic saddle point being $c(\vs) = 1/2^n$.
The free energy density $\beta f_N(\beta) = -(1/N)\expect\, \log Z_N$
is then obtained  by taking the $n\to 0$ limit:
\begin{eqnarray}
\beta f_{N}(\beta) & = &-\log 2-\frac{\gamma}{2}\expect\, \log\cosh\beta J +
\frac{1}{N}\,\beta f^{(1)} +O(N^{-2})\, ,\label{eq:Replica1}\\
\beta f^{(1)} &= & -\frac{1}{2}\,\gamma^2\, \expect\log\cosh\beta J
+\frac{1}{4}\gamma^2\, \expect\log\cosh\beta(J_1+J_2)\label{eq:Replica2}\\
&&-\frac{1}{2}\sum_{k=1}^{\infty}\frac{\gamma^k}{k}\,\expect
\log\big\{1+\prod_{i=1}^k\tanh\beta J_i\big\} .\nonumber
\end{eqnarray}
Here expectations are taken with respect to the $J_1,J_2,\dots$ which are
i.i.d. with distribution $p(J)$.
The $O(1/N)$ term is the contribution of Gaussian fluctuations around
this saddle point. As expected, it diverges upon approaching 
the spin glass critical temperature $\beta_{\rm c}$
(defined by $\gamma\,\expect\, \tanh^2\beta_{\rm c}J = 1$).
More precisely we have
\begin{eqnarray}
f^{(1)}(\beta) = -\frac{1}{4}\log(\beta_{\rm c}-\beta) + O(1)\, .
\end{eqnarray}

We shall now rederive the above results by using the approach outlined
in the previous Section. A serious shortcoming of this approach is that 
it does not provide an explicit expression for the free energy.
One can circumvent this problem by considering the 
internal energy density $u_N(\beta) = \expect\, E(\sigma)/N$.
Since all the edges of the graph are equivalent,
\begin{eqnarray}
u_N(\beta) = -\frac{N-1}{2N}\, \gamma\,\expect\left\{ 
J_{ij}\<\sigma_i\sigma_j\>\,|\, (i,j)\in \G\right\}\, .\label{eq:FirstEnergy}
\end{eqnarray}
Here $\expect\left\{ \; \cdot\;|\, (i,j)\in G\right\}$ denotes expectation 
conditional to the edge $(i,j)$ belonging to the graph $\G$.
Consider now a particular graph $\G$ in which the link $(i,j)$ is present.
It is simple to show that
\begin{eqnarray}
\<\sigma_i\sigma_{j}\> =
t_{ij} +(1-t_{ij}^2)\, \frac{C^{(ij)}_{ij}}{1+t_{ij}C^{(ij)}_{ij}}\, ,
\end{eqnarray}
where $C^{(ij)}_{ij}$ denotes the correlation $\<\sigma_i\sigma_j\>$
after link $(i,j)$ has been removed (notice that local magnetizations 
$\<\sigma_i\>$ vanish by symmetry), and $t_{ij} = \tanh\beta J_{ij}$.
Notice that sampling $\G$ under the condition $(i,j)\in\G$,
and then removing $(i,j)$ is equivalent (fro the Erd\"os-Renyi ensemble)
to sampling $\G$ under the condition $(i,j)\not\in\G$.
Substituting in Eq.~(\ref{eq:FirstEnergy}), we get
\begin{eqnarray}
u_N(\beta) = -\frac{N-1}{2N}\, \gamma\,\expect\{Jt_J\}
-\frac{N-1}{2N}\, \gamma\,\expect\left\{\left.J(1-t^2_J)
\frac{C_{ij}}{1+t_J\,  C_{ij}}\, 
\right|\,  (i,j)\not\in \G\right\}\, ,\phantom{A}
\end{eqnarray}
where we used the shorthand $t_J=\tanh\beta J$. The second term vanishes 
as $N\to\infty$, since it behaves as the correlation between two 
uniformly random sites in the system. We will therefore estimate it to the 
leading non-trivial order. Moreover, we can expand $(1+t_J\,  C_{ij})^{-1}$
in an absolutely convergent series to get
\begin{multline}
u_N(\beta) = -\frac{\gamma}{2}\,\expect\{Jt_J\}+\frac{1}{N}
\frac{\gamma}{2}\,\expect\{Jt_J\}-\\
-\frac{\gamma}{2}\, \sum_{k=0}^{\infty}\,(-1)^k\,
\expect\{ J(1-t^2_J) t_J^k\}\;
\expect\{C_{ij}^{k+1}|\,  (i,j)\not\in \G\} +O(N^{-2})\, ,\phantom{A}
\label{eq:EnergyExpansion}
\end{multline}
where we factorized the expectation thanks to the fact that $\G$
does not contain $(i,j)$ and therefore $C_{ij}$ is independent of $J$.

We are left with the task of computing the moments of $C_{ij}$. According 
to our general strategy, we use the identity
$C_{ij} = \frac{1}{\beta}\left.\frac{\partial M_i}{\partial H_j} \right|_{H=0}$.
It is well known~\cite{MezardParisiBethe} that, for the Erd\"os-Renyi random 
graph, 
$M_l\ed \tanh\beta h_{i}^{(j)}$ up to corrections vanishing as $N\to\infty$
(here $\ed$ denoted equality in distribution).  Furthermore, 
by differentiating Bethe equations (\ref{eq:BetheEquations}) we get
\begin{eqnarray}
\frac{\partial h^{(m)}_i}{\partial H_j} 
= \delta_{ij}+\sum_{l\in\partial i\backslash m}
\tanh\beta J_{il}\,  \frac{\partial h^{(i)}_l}{\partial H_j} \, .
\label{eq:ResponseGraph}
\end{eqnarray}
By averaging over the graph and couplings and recalling that 
the degree of a site is,  for large $N$, a Poisson random 
variable of mean $\gamma$, we obtain (here we use the shorthand 
$\partial_j$ for partial derivative with respect to $H_j$):
\begin{eqnarray}
\expect\left\{\left(\partial_j h^{(m)}_i\right)^k\right\}
& = &\frac{1}{N} + \gamma \,\expect\{t_J^k\} \,
\expect\left\{\left(\partial_j h^{(i)}_l\right)^k\right\}
+O(N^{-2})\, .
\end{eqnarray}
Moreover if we condition on $i$, $j$ being distinct and not connected
by an edge, the term $\delta_{ij}$ in Eq.~(\ref{eq:ResponseGraph})
is surely missing for at least two iterations, leading to
\begin{eqnarray}
\expect\left\{\left.\left(\partial_j h^{(m)}_i\right)^k\right|
i\neq j,\, (i,j)\not\in\G\right\}
& = & (\gamma \,\expect\{t_J^k\})^2 \,
\expect\left\{\left(\partial_j h^{(i)}_l\right)^k\right\}+O(N^{-2})\, .
\phantom{A}
\end{eqnarray}
By solving these equations and identifying 
$\expect\{C_{ij}^{k+1}|\,  (i,j)\not\in \G\}$ with the left hand side of the 
last equation, we finally get
\begin{eqnarray}
\expect\{C_{ij}^{k}|\,  (i,j)\not\in \G\} = \frac{1}{N}
\,\frac{(\gamma \,\expect\{t_J^k\})^2}{1-\gamma \,\expect\{t_J^k\}} +
O(N^{-2})\, ,\label{eq:CorrelationResult}
\end{eqnarray}
for $k$ even (for $k$ odd the expectation vanishes by symmetry. 
We can now plug this into Eq.~(\ref{eq:EnergyExpansion}) to get the 
final result
\begin{eqnarray}
u_N(\beta)&=& -\frac{\gamma}{2}\,\expect\{Jt_J\}+\frac{1}{N} \, u^{(1)}(\beta)
+O(N^{-2})\, ,\\
u^{(1)}(\beta) & = & \frac{\gamma}{2}\,\expect\{Jt_J\}
+\frac{\gamma}{2}\, \sum_{k\, \, \mbox{\tiny odd}} \, 
\expect\{ J(1-t^2_J) t_J^k\}\, 
\frac{(\gamma \,\expect\{t_J^{k+1}\})^2}{1-\gamma \,\expect\{t_J^{k+1}\}}\, .
\end{eqnarray}
One can then compute the free energy density by integrating over the 
temperature with boundary condition $\beta f_{N}(\beta)\to -\log 2$
as $\beta\to 0$. It is easy to check
that the resulting expression
coincides with the replica result (\ref{eq:Replica1}),
(\ref{eq:Replica2}).

One surprising feature of the above calculation is 
the behavior of correlations. One would naively assume that 
the correlation between two random spins is concentrated around
a typical value of order $N^{-\delta}$, with $\delta>0$, leading to
$\expect\{C_{ij}^k\}\sim N^{-k\delta}$, in contradiction with the correct 
result, Eq.~(\ref{eq:CorrelationResult}). It is therefore interesting
to study the distribution of $C_{ij}$. For the sake of simplicity we
shall  admit the cases $i=j$ and $(i,j)\in\G$ which where excluded in 
the above calculation. In the $N\to\infty$ limit, the correlations
satisfy the same distributional equation as the response functions,
cf.~Eq~(\ref{eq:ResponseGraph}):
\begin{eqnarray}
C \ed \sum_{i=1}^{k} (\tanh\beta J_i)\, C_i\, .
\end{eqnarray}
Here $k$ is a Poisson random variable of mean $\gamma$ and $J_i$
are i.i.d. with distribution $P(J)$. Notice that this equation 
can only fix the distribution of $C$, to be denoted by $\rho(C)$,
up to a scaling factor. In fact, if $C$ is a random variable satisfying the 
above equation, also $a\, C$  does. We shall therefore 
write $\rho(C) = (1/C_0)\rho_*(C/C_0)$, where the solution $\rho_*(C)$
is fixed arbitrarily and $C_0=C_0(N)$ is the typical scale of
correlations in a system of size $N$. 
The scale $C_0$ will be determined by a matching procedure.

Consider the characteristic function 
$\phi(s) \equiv \int \exp(isC)\, \de\rho_*(C)$. This satisfies the equation
\begin{eqnarray}
\phi(s) = \exp\{-\gamma[1-\expect\, \phi(ts)]\}\, ,
\end{eqnarray}
where expectation is taken with respect to $t = \tanh\beta J$.
From this is easy to derive the small $s$ behavior 
$\phi(s) \simeq 1- \phi_0|s|^{\alpha}$ where we can fix the freedom
in the scale of  $C$ by setting $\phi_0=1$. The exponent $\alpha$
is determined by
\begin{eqnarray}
\gamma\, \expect\{|\tanh\beta J|^{\alpha}\} = 1\, .
\end{eqnarray}
Therefore $\alpha$ grows monotonously with $\beta$ and takes the value 
$\alpha=2$ at $\beta = \beta_{\rm c}$. The large $C$ behavior of the 
correlations distribution is $\rho_*(C)
\simeq (\frac{1}{\pi}\Gamma(1+\alpha)\sin(\frac{\pi\alpha}{2}))|C|^{-1-\alpha}$.
The physical reason of this power law 
tail is easily understood. Consider the neighborhood of a site $i$.
Asymptotically, this will be a random tree with Poisson degree distribution
(a Galton-Watson tree). For a site $j$ at distance $r$ from $i$, 
we have $C_{ij} = \tanh\beta J_1\cdots \tanh\beta J_r$, where
$J_1,\dots,J_r$ are the couplings on the path joining $i$ to $j$.
If we only consider these sites and sum over all finite $r$
as $N\to\infty$, we get
\begin{eqnarray}
\rho_{\rm neigh}(C) = \frac{1}{N}\sum_{r=0}^{\infty}\gamma^{r}\,
\expect\,\,\delta\!\left(C-\prod_{i=1}^r\tanh\beta J_i\right)\, .
\end{eqnarray}
The small $C$ asymptotics of this distribution is $\rho^{\rm neigh}(C)
\simeq\rho_0^{\rm neigh} \, |C|^{-1-\alpha}$. By computing
$\rho_0^{\rm neigh}$ and matching this behavior with the large 
$C$ behavior of $\rho(C) = (1/C_0)\rho_*(C/C_0)$, we determine
\begin{eqnarray}
C_0(N) = \left\{\frac{\pi}{\Gamma(1+\alpha)\sin\frac{\pi\alpha}{2}}\,
\frac{1}{(-\gamma\expect\, |t_J|^{\alpha}\log|t_J|)}
\right\}^{1/\alpha}\, N^{-1/\alpha}\, ,
\end{eqnarray}
where $t_J=\tanh\beta J$.
As the temperature decreases from $\infty$ to the critical temperature,
$\alpha$ increases from $0$ to $2$ and therefore the typical correlation scale
increases from $N^{-\infty}$ to $N^{-1/2}$. However, correlations
are never concentrated around a particular value but have a power-law
behavior at all temperature. Integer moments are therefore governed by the 
largest correlations in the system (in particular they are ruled by 
$\rho^{\rm neigh}(C)$) and are always of order $N^{-1}$.
Finally notice that, because of the power-law behavior, there is no 
definite loop length responsible for corrections to BA.

Let us conclude with a comment. One could have been skeptical
about the success of the present approach in computing 
$1/N$ effects in for spin models on random graphs. In fact, an average 
fraction $1/N$ of spins of such systems lies in a neighborhood of finite-size
loops (e.g. triangles). For such spins the violation of BA is 
non-perturbative and our approach could have seemed a priori 
hopeless. However, the exactness of our method for uni-cyclic  graph 
allows to overcome this problem. On the other hand, it was crucial not to 
neglect terms of order $\expect\,\{C^{k+1}_{ij}\}$, $k>0$ in 
Eq.~(\ref{eq:EnergyExpansion}), i.e. to follow the procedure 2
described in  Section \ref{se:GeneralSection}.

The same kind of argument suggest that the systematic expansion 
described in Section \ref{se:GeneralSection} corresponds in fact to 
the $1/N$ expansion for random graphs. The next correction is due to couples 
of joined closed loops, an event occurring in average near a fraction 
$1/N^2$ of the sites.
%
%************************************************************
%
\section*{Acknowledgments}

While this paper was being completed, we discovered that Giorgio Parisi
and Franti\v{s}ek Slanina had been working at a similar loop expansion
using a completely different approach~\cite{ParisiSlanina}.
We are grateful to them for letting us know about their results.

This work was supported by EVERGROW, i.p. No. 1935 in the
complex systems initiative of the Future and Emerging Technologies
directorate of the IST Priority, EU Sixth Framework.
%
%************************************************************
%


\begin{thebibliography}{99}

\bibitem{Bethe} H.~A.~Bethe, Proc.~Roy.~Soc.~of London~A~{\bf 150}, 552 (1935) 

\bibitem{Jordan} M.~I.~Jordan.  Statistical  Science
(Special Issue on Bayesian Statistics), {\bf 19}, 140 (2004)

\bibitem{Pearl} J.~Pearl. {\it Probabilistic reasoning in intelligent 
systems: networks of plausible inference}. Morgan Kaufmann, 1988

\bibitem{Yedidia} J.~S.~Yedidia, W.~T.~Freeman, and Y.~Weiss, 
{\em ``Constructing Free Energy Approximations and 
Generalized Belief Propagation Algorithms''}. 
submitted to IEEE Trans.~Info.~Theory. {\tt MERL TR 2002-35}

\bibitem{RichardsonUrbanke} T.~Richardson and R.~Urbanke,
in {\it Codes, Systems, and Graphical Models}, edited by
B.~Marcus and J.~Rosenthal, Springer, New York, 2001.

\bibitem{MarcGiorgioRiccardo} M.~M\'ezard, G.~Parisi, and R.~Zecchina,
Science {\bf 297}, 812 (2002).

\bibitem{Kikuchi} R.~Kikuchi. Phys.~Rev.~81, 988 (1951)

\bibitem{GeorgesYedidia} A.~Georges and J.~Yedidia, 
J.~Phys.~A:~{\bf 24} 2173

\bibitem{ZinnJustin} J.~Zinn-Justin, 
{\it Quantum Field Theory and critical Phenomena},
Oxford University Press, 1989

\bibitem{Cardy} J.~L.~Cardy, {\it Scaling and Renormalization in Statistical 
Physics}, Cambridge University Press, 1996

\bibitem{MezardParisiBethe} M.~M\'ezard and G.~Parisi,
Eur.~Phys.~J.~B~{\bf 20}, 217 (2001)

\bibitem{Yedidia2} J.~S.~Yedidia, W.~T.~Freeman, Y.~Weiss, in
{\it Exploring Artificial Intelligence in the New Millennium}, G.~Lakemeyer
ed., Morgan  Kaufmann, 2003

\bibitem{Barber} M.~N.~Barber, R.~B.~Pearson, D.~Toussaint, and 
J.~L.~Richardson, 
Phys.~Rev.~B~{\bf 32}, 1720-1730 (1985)

\bibitem{KimPatrascioiu} J.-K.~Kim and A.~Patrascioiu,
Phys.~Rev.~D~{\bf 47}, 2588-2590 (1993)

\bibitem{ParisiRuiz}  G.~Parisi and J.~J.~Ruiz-Lorenzo,
Phys.~Rev.~B~{\bf 54}, R3698-R3701 (1996)

\bibitem{Gaunt} D.~S.~Gaunt, M.~F.~Sykes, and S.~McKenzie, J.~Phys.~A 
{\bf 12}, 871 (1979)

\bibitem{VianaBray} L.~Viana and A.~Bray, J.~Phys.~C {\bf 18} 3037

\bibitem{GuerraToninelli} F.~Guerra, F.~L.~Toninelli J.~Stat.~Phys, 
531, 115 (2004)

\bibitem{Campellone} M.~Campellone, G.~Parisi, and M.~A.~Virasoro,
{\em ``Replica method and finite volume corrections''}. To be published.

\bibitem{MonassonDiluted} R.~Monasson, J.~Phys.~A {\bf 31} 513 (1998)

\bibitem{ParisiSlanina} G.~Parisi, F.~Slanina,
{\em ``Loop expansion around the Bethe Peierls approximation for
lattice models''}, in preparation.

\end{thebibliography}
\end{document}